\documentclass[a4paper,onecolumn,accepted=2024-12-27]{quantumarticle}
\pdfoutput=1
\usepackage[utf8]{inputenc}
\usepackage[english]{babel}
\usepackage[T1]{fontenc}
\usepackage{amsmath}
\usepackage{hyperref}

\usepackage{amsfonts, amsmath, bm, bbm, graphicx, epsfig, epstopdf}
\usepackage{dcolumn}
\usepackage{textcomp}
\usepackage{soul}
\usepackage{graphicx}
\usepackage{amsfonts, amsmath, bm, bbm, graphicx, epsfig, epstopdf}
\usepackage{dcolumn}
\usepackage{textcomp}
 \usepackage[caption=false]{subfig}
\usepackage{soul}
\usepackage{tikz}
\usepackage{lipsum}

    \usepackage{braket}
\renewcommand\bra[1]{{\langle{#1}|}}
\makeatletter
\renewcommand\ket[1]{%
  \@ifnextchar\bra{\k@t{#1}\!}{\k@t{#1}}%
}
\newcommand\k@t[1]{{|{#1}\rangle}}
\makeatother

\begin{document}

\title{ Ultratight confinement of atoms in a Rydberg empowered optical lattice}
\author{Mohammadsadegh Khazali}
\affiliation{Department of Physics, University of Tehran, Tehran 14395-547, Iran}
\affiliation{Email: mskhazali@ut.ac.ir}
\orcid{0000-0002-7244-3543}

\begin{abstract}
Optical lattices serve as fundamental building blocks for atomic quantum technology. However, the scale and resolution of these lattices are diffraction-limited to the light wavelength. In conventional lattices, achieving tight confinement of single sites requires high laser intensity, which unfortunately leads to reduced coherence due to increased scattering. 
This article presents a novel approach for creating an atomic optical lattice with a sub-wavelength spatial structure. 
The potential is generated by leveraging the nonlinear optical response of three-level Rydberg-dressed atoms, which allows us to overcome the diffraction limit of the driving fields. 
The resulting lattice comprises a three-dimensional array of ultra-narrow Lorentzian wells over nanometer scales. 
These unprecedented scales can now be accessed through a hybrid scheme that combines the dipolar interaction and optical twist of atomic eigenstates.
 The interaction-induced two-body resonance that forms the trapping potential, only occurs at a peculiar laser intensity, localizing the trap sites to ultra-narrow regions over the standing-wave driving field. 
The feasibility study shows that single-atom confinement in Lorentzian sites with 3nm width, and 37MHz depth are realizable with available lasers. 
The development of these ultra-narrow trapping techniques holds great promise for applications such as Rydberg-Fermi gates, atomtronics, quantum walks, Hubbard models, and neutral-atom quantum simulation.
\end{abstract}

\maketitle

\section{Introduction}
The primary enabling technology in atomic quantum processors is the coherent control of the position and motion of atoms by lasers.
The underlying mechanism in conventional optical lattices is the ac-stark shift of atomic levels formed by far-off-resonant laser fields.
The diffraction limit, which is about the wavelength of the light, is what determines the scale and spatial resolution of such optical potential landscapes. 
This fundamentally limits the optical manipulation of atoms, affecting some of the quantum technology applications.
  For instance, in the recently proposed Rydberg-Fermi quantum simulator \cite{Kaz21,Khaz22Log}, ultra-tight confinement of atoms within the single lobe of the Rydberg wave function is required for high-fidelity scalable quantum processing. Tight confinement is also demanding for applications based on distance selective interaction \cite{Kha22, Hol22} and controlled Rydberg anti-blockade operations \cite{Ate07,Wu22}. Finally, tight confinement is demanded to improve the fidelity of neutral atom processors \cite{Gra19,Pag22,Cet22,Bre99}.

   \begin{figure} 
\centering 
       \scalebox{0.55}{\includegraphics{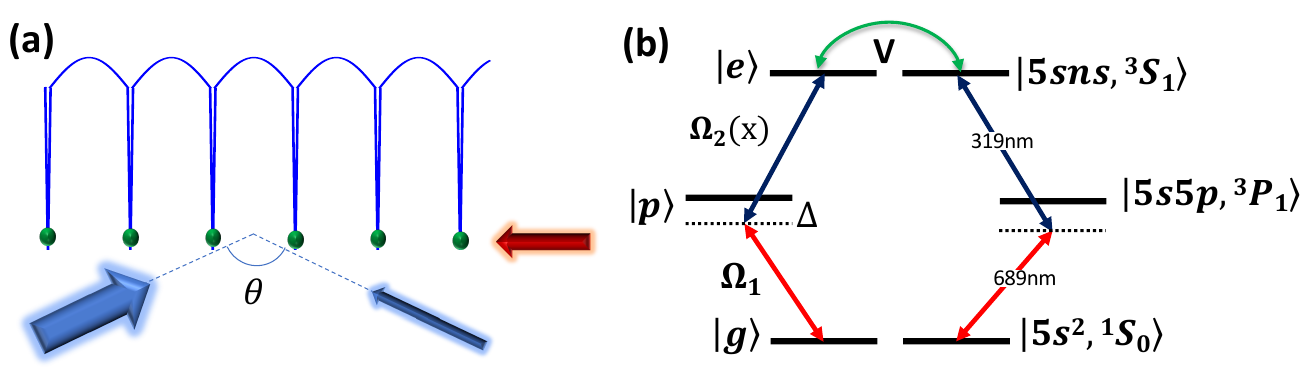}} 
\caption{Ultra-tight confinement of atoms in an interaction-induced atomic lattice. (a) Rydberg dressing of ground state Sr atoms with standing-wave laser field $\Omega_2(x)$ results in a combination of classical AC stark shift potential that follows the standing wave profile in addition to interaction-induced periodic trapping potential that features sharp resonance at the anti-nodes of the standing wave.
The ultra-narrow trapping wells facilitate ultra-tight atom confinement. (b) The level scheme presents in-resonance two-photon Rydberg dressing. }\label{Fig1}
\end{figure}

 Tight confinement of atoms in conventional optical lattices requires extensive power i.e. the spatial width of the ground motional state is inversely proportional to the quadruple root of the laser intensity.   
 The drawback is the loss of coherence due to the enhanced scattering.
In an alternative approach, this article deploys the nonlinear response of Rydberg-dressed atoms to the intensity of a standing-wave driving field, as a means to form a lattice of ultra-narrow trapping potentials. 
The sub-wavelength resolution arises when the composition of eigenstates on a two-atom basis twists rapidly at a specific light intensity to form interaction-induced resonance over a short length scale of the standing wave.  Unlike the conventional ac-stark shift potentials, this interaction-induced potential is a quantum effect, with magnitude proportional to $\hbar$.
This effect forms  3D lattices with potential widths as small as the Bohr radius. 
Furthermore, the spatial correlation of two interacting atoms in a standing wave driving field forms a phononic bus that facilitates all-to-all connectivity for gate operations via the side-band laser excitations similar to the ion trap platforms \cite{Cir95}. 

Previous related work, such as Refs. \cite{Wan18, Tsu20}, explored creating lattices of sub-wavelength repulsive barriers using a non-interaction-based dark-state model. The dark state model could not generate attractive potentials and only forms box-type potentials, not applicable for tight confinement of atoms.
 In contrast, the present work leverages Rydberg interactions to create a lattice of ultra-tight trapping potentials at the positions of interaction-induced resonances. Additionally, the Rydberg interaction facilitates a phononic bus, with potential applications in quantum information processing.

The recent advances in optical control of Rydberg atoms have opened a wide range of applications in quantum technology \cite{Saf10,Kha20,Kha19,Kha15,Khaz22AllOptic,Kha17,KazRev}.
The required dipolar interaction in this proposal is formed by dressing ground-state atoms with the highly excited Rydberg state \cite{Zei16,Hin23,Eck23,Kha16,Kha18,Khazal24,San00,Hon10} in a resonant driving scheme \cite{Gau16,Kha21}.
Rydberg dressing of a BEC with homogeneous laser lights could form triangular and quasi-ordered droplet crystals \cite{Kha21,Hen12,Shi21,Shi20,ZShi21,Shi23,ZShi23,ZeyShi23,Shi22,Shi24,ZShi24,ZeyShi24}. However, this periodic structure would not be fixed in space. 
In contrast, in this proposal, the spatial pattern of the driving field and intensity dependence of the potential would spatially pin the lattice sites to the nodes of the standing wave. Therefore, the lattice structure would be fixed in the space. This feature is required to address individual sites in atomic processors. 

\section{Scheme}

The atomic lattice {\it scheme} is based on dressing  $^{88}Sr$ atoms with the highly excited Rydberg level, see Fig.~\ref{Fig1}.
In the two-photon in-resonance dressing scheme \cite{Gau16,Kha21}, the single atom Hamiltonian is given by
\begin{equation}
{H}_i/\hbar=\frac{\Omega_1}{2} ({\sigma}^i_{gp}+{\sigma}^i_{pg})+\frac{\Omega_2(x_i)}{2}  ({\sigma}^i_{ep}+{\sigma}^i_{pe})-\Delta {\sigma}^i_{pp},
\end{equation} 
where $\sigma^i_{\alpha,\beta}=|\alpha \rangle\langle \beta|$ is the transition operator of $i^{th}$ atom. The two Rabi frequencies $\Omega_{1,2}$ are applied by 689nm and 318nm lasers that are detuned from the intermediate state $\ket{p}$ by $\Delta$.
With negligible Rydberg decay rates, the system would follow the dark eigen-state $| d \rangle \propto \Omega_2 | g \rangle - \Omega_1 | e \rangle$ with zero light-shift. In the $\Omega_1\ll \Omega_2$ limit, ground state atoms will be partially dressed by Rydberg states with the population of $P_e=(\Omega_1/\Omega_2)^2$.  This small Rydberg population would get polarized and trapped by the standing-wave  $\Omega_2(x)$ laser via the AC stark shift. This shallow sinusoidal trap guides atoms to the standing wave anti-nodes where they experience the interaction-induced quantum trap discussed below, see Fig.~1a.

 \begin{figure} [h!]
\centering 
       \scalebox{.7}{\includegraphics{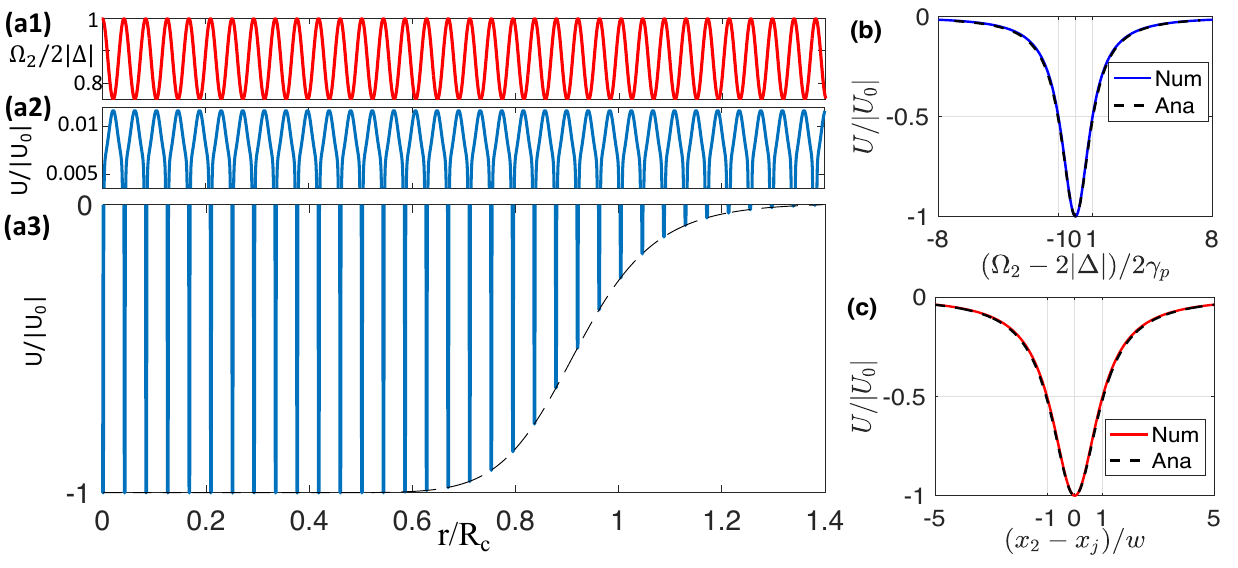}} 
\caption{Interaction-induced trap.  In a simplified picture the first atom is located at the origin, fixed at the anti-node of the standing wave.  (a1) The red line shows the spatial profile of the Rabi frequency $\Omega_2(x)=\Omega_{2c}+\Omega_{2sw}|\sin(kx\sin(\theta/2))|$, where $k=2\pi/\lambda$. (a2-a3) The blue line shows the interaction of two atoms as a function of interatomic distance. (a2) A shallow ac-stark shift-induced potential guides the atoms toward the antinodes where (a3) the interaction-induced resonance forms deep ultra-narrow quantum trapping. 
With a single atom per lattice site, the effective trapping interaction would be the sum of two-body interactions of all the sites within the $\pm R_c$ distance. 
(b) Assuming the first atom to be at the origin, the interaction-induced resonance occurs at the nodes of standing-wave $\Omega_2(x_2)=2|\Delta|$, with the maximum depth of $U_{0}=\frac{\hbar\Omega_1^4}{8\Delta\gamma^2}$ and the HWHM of $\Omega_2(x_2)-2|\Delta|=\pm 2 \gamma_p$. The analytical form of Eq.~\ref{Eq_5} and numerical calculation of the interaction potential  (Eq.~\ref{Eq3}) presents a perfect match.  
(c) Spatial form of the interaction-induced trap for the second atom at the position of the $j^{th}$ standing wave node. The Lorentzian potential of Eq.~\ref{Eq_Lorentzian} with the width $w$ (Eq.~\ref{Eq_Width}) shows a perfect match with the numeric calculation of Eq.~\ref{Eq3}.
Chosen parameters in (a) are $\Omega_{2c}=2|\Delta|=2\pi\times10$MHz,  $\Omega_{2sw}=|\Delta|/2$, the maximum loss rate per atom in a pair interaction is limited to 1Hz, $n=100$, $\theta=\pi$. 
 }\label{Fig2}
\end{figure} 

The van-der Waals interaction between Rydberg atoms ${V}_{ij}=\hbar C_6/r_{ij}^6 {\sigma}_{ee}^i {\sigma}_{ee}^j$ is  a function of interatomic distance $r_{ij}$. 
The dynamic of the system under Rydberg interaction is governed by the master equation of two-body density matrices. The two-body density matrices ${\rho}_{ij}=\text{Tr}_{\bar{i},\bar{j}}{\rho}$  are obtained by tracing over all except $i$ and $j$ atoms. The corresponding master equation would be given by
\begin{equation}\label{Heisenberg}
\partial_t {\rho}_{ij}=-\frac{\text{i}}{\hbar}[{H}_i+{H}_j+{V}_{ij},{\rho}_{ij}] + \mathcal{L}_i({\rho}_{ij})+\mathcal{L}_j({\rho}_{ij}) 
\end{equation}
The internal state dynamics are governed by single-particle dissipation described by  $\mathcal{L}_i$ operator acting on  $i$th atom. 
The Liouvillian term $\mathcal{L}_i (\rho)=\sum_\beta \mathcal{D}(c_{\beta})\rho_i$ with $\mathcal{D}(c)\rho_i=c\rho_ic^{\dagger}-1/2 (c^{\dagger}c \rho_i+\rho_ic^{\dagger}c )$ in
the Lindblad form governs the dissipative time evolution. Lindblad terms encounter spontaneous emission from Rydberg $c_{pe}=\sqrt{\gamma_e }|p\rangle \langle e|$  and intermediate level $c_{gp}=\sqrt{\gamma_p}|g\rangle \langle p|$. The  spontaneous emission rates are $\gamma_{p}/2\pi=7.6$kHz  and $\gamma_{e}$ can be found in \cite{Kun93}.

Considering the steady state $\bar{\rho}_{ij}$ of Eq.~\ref{Heisenberg},  the effective interaction  would be given by  
\begin{equation}
\label{Eq3}
{U}(r_{ij})=\text{Tr}[\bar{\rho}_{ij} ({H}_i+{H}_j+{V}_{ij})].
\end{equation}
 For homogeneous lasers, a plateau-type interaction profile would be formed with constant interaction within the soft-core as depicted by the dotted line in Fig.~\ref{Fig2}a.
 In  Rydberg-dressing the interaction region is defined by  $R_c$; the interatomic distance within which interaction-induced detuning equals the effective power broadening $P_eV(R_c)=\Omega_1\Omega_2/2\Delta$ \cite{Kha21}. 
 The soft-core interaction features a sharp peak at $\Omega_2=2|\Delta|$, due to an interaction-induced resonance, see Fig.~\ref{Fig2}b.

To form the optical lattice with the mentioned interaction-induced resonance, a space-dependent variation of the upper laser is deployed.
Using different intensities for the counter-propagating 318nm lasers, results in the desired spatial pattern of the Rabi frequency  
\begin{equation}
\label{Eq_Om2}
\Omega_2(x)=\Omega_{2c}+\Omega_{2sw}|\sin(k_2x\sin(\theta/2))|,
\end{equation}
 where $k_2=2\pi/\lambda_2$ is the  wave-vector of $\Omega_2$ laser and $\theta$ is the angle between counter propagating lights, see Fig.~\ref{Fig1}a. In the following, we consider counter-propagating lights $\theta=\pi$.
The standing wave first of all forms a shallow AC stark shift optical lattice by polarizing the minor Rydberg population as $U_{AC}=P_e\Omega_{2sw}^2\sin^2(k_2x\sin(\theta/2))/8\Delta$ which guides the atoms to the nodes of standing wave.  At this point $\Omega_{2max}=2|\Delta|$ provides resonant interaction upon the presence of at least two atoms within the interacting region $R_c$. This resonant interaction forms the quantum trap.

To give a simplified picture of interaction-induced trapping potential, we initially assume that one atom is fixed at the origin. The trapping potential at the position of the second atom $x$ is plotted in Fig.~\ref{Fig2}a,c. 
 Considering the spatial variation of the $\Omega_2$ in Eq.~\ref{Eq_Om2} over the narrow area $k_2.(x-x_j)\ll1$ around the $j^{th}$ node of the standing wave, the trapping profile could be approximated by a Lorentzian function 
\begin{equation}
\label{Eq_Lorentzian}
U^q=\frac{U^q_0}{1+(x-x_j)^2/w^2}
\end{equation}
where the half-width at half-maximum  and the depth of the spatial trap well are given by
\begin{equation}
\label{Eq_Width}
 w=\frac{2\gamma_p}{k_2\sin(\theta/2)\Omega_{2sw}}; \quad \quad   \quad \quad  U^q_{0}=\frac{\hbar\Omega_1^4}{8\Delta\gamma^2}. \quad \quad    \quad  
\end{equation}
Figure \ref{Fig2}c compares this analytical form of Eq.~\ref{Eq_Lorentzian} with the numerical results obtained from Eq.~\ref{Eq3}.
The scale of the trap width as a function of  $\Omega_{2sw}$ is plotted in Fig.~\ref{Fig5}c. Remarkably, with   $\Omega_{2sw}/2\pi=1.7$MHz the trap width would be as tight as the radius of $^{88}$Sr atoms.

 The trapping potential in Fig.~\ref{Fig2}a contains an attractive well for overlapping atoms at the origin. However, the collisional blockade mechanism ensures that only one atom is loaded per lattice site within the small trapping volume \cite{Sch01,Sch02}. The effective potential experienced by each site is determined by the two-body interactions of neighboring lattice sites within the interacting distance $R_c$. By conducting numerical simulations with three atoms localized within the interaction core $R_c$, it is found that the total potential of the system is the sum of the binary interactions, namely $U_{12}+U_{23}+U_{13}$. The interaction-induced loss rate also follows the interaction pattern and results to $\Gamma_{12}+\Gamma_{23}+\Gamma_{13}$.  However, in the case of a large spacing where $R_{12}, R_{23}<R_c<R_{13}$, the total potential would be twice the binary case, i.e., $U_{12}+U_{23}$ with the interaction induced loss rate of $\Gamma_{12}+\Gamma_{23}$. Hence, increasing the interaction region speeds up the loading process while preserving the interaction-to-loss ratio.
The latter case explains a 1D lattice with the blockade radius only covering the next nearest neighbors. Since no more than one atom can be trapped within the same lattice site, this 1D lattice will fulfill the assumption $R_{i,i+1},R_{i,i-1}<R_c<R_{i-1,i+1}$ where $i$ span over all lattice sites.
Finally, it should be clarified that the lattice size is determined by the size of the standing wave $\Omega_2$ and not limited to the size of interaction region  $R_c$. However, the depth of each ultra-tight quantum well is determined by the number of filled sites within the $R_c$ distance.

\begin{figure} 
\centering 
       \scalebox{0.7}{\includegraphics{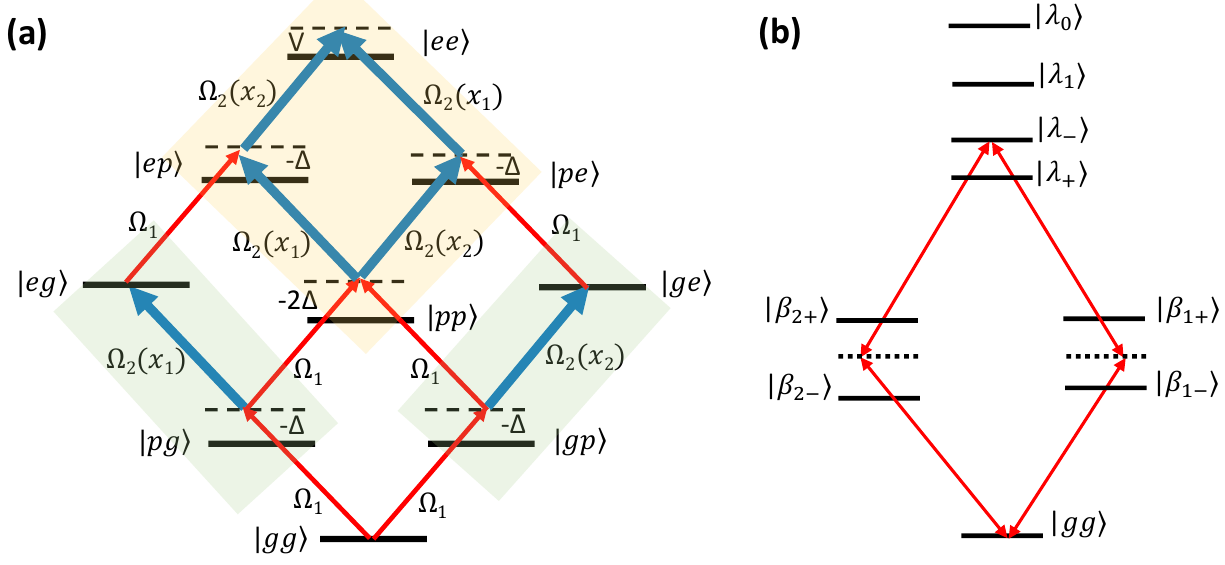}} 
\caption{ The origin of the trapping potential is the interaction-induced resonance.  
(a) With $\Omega_1\ll\Omega_2$ the two-atom Hilbert space would be organized in three subspaces of ground state, single-excitation (green boxes) and double-excitations (yellow box) that are coupled by weak $\Omega_1$ laser. (b) The effects of strong coupling $\Omega_2$ and interaction $V$ could be observed by diagonalizing the green and yellow subspaces with eigen-states of $|\beta_{i\pm}\rangle$ and $|\lambda_{0,1,\pm}\rangle$ respectively.  The interaction-induced level shift, makes the $|\lambda_{-}\rangle$ in-resonance with the ground state at positions with $\Omega_2(x_1)^2+\Omega_2(x_2)^2=8\Delta^2$. This resonance significantly enhances the interaction at a very precise position in $\Omega_2$ standing wave and thus forms ultra-narrow trapping potential. 
 }\label{Fig3}
\end{figure}

A more accurate form of the interaction encounters the motion of two interacting atoms in the lattice. This requires encountering distinguished space-dependent Rabi frequencies $\Omega_2(x_1)$ and $\Omega_2(x_2)$ for two atoms.  
Considering the level scheme of Fig.~\ref{Fig3}a in a two-atom basis, the doubly excited Rydberg state asymptotically decouples within the interaction region $R_c$ as  $V\rightarrow \infty$. Taking into account the remaining states, in the limit of $\Omega_1\ll\Omega_{2c}$ the steady state density could be obtained by adding three orders of perturbative corrections to the initial ground state. In the limit of $\gamma_p\ll\Delta$ the dressing interaction between two atoms located at $x_1$ and $x_2$ inside the interaction region $x_1-x_2<R_c$   is given by 
\begin{widetext}
\begin{equation}
\label{Eq_5}
U^q(x_1,x_2)=\frac{\hbar\Omega_1^4}{4\Omega_2^2(x_1)\Omega_2^2(x_2)\Delta}
\frac{[8\Delta^2(\Omega_2^2(x_1)+\Omega_2^2(x_2))-(\Omega_2^2(x_1)-\Omega_2^2(x_2))^2](\Omega_2^2(x_1)+\Omega_2^2(x_2)+8\Delta^2)}{(\Omega_2^2(x_1)+\Omega_2^2(x_2)-8\Delta^2)^2+64\gamma_p^2\Delta^2}.
\end{equation}
\end{widetext}
The maximum interaction occurs when the denominator is minimized at $\Omega_2(x_1)^2+\Omega_2(x_2)^2=8\Delta^2$. 
Note that the attractive or repulsive nature of the potential peak is determined by the sign of detuning $\Delta$.
The presented analytic model of Eq.~\ref{Eq_5} perfectly resembles the numerical results, see  Fig.~\ref{Fig2}, \ref{FigMeanfield}. 
Figure ~\ref{FigMeanfield}d, plots the ground state wave function of two atoms under dressing trap potential of Eq.~\ref{Eq_5} as a function of the first and second atoms' positions $x_1$ and $x_2$. The mean field approach used the imaginary time evolution of the Gross Pitaevski Equation for calculating the ground motional state in Fig.~\ref{FigMeanfield}d.

\section{The Interaction-Induced Resonance }
\begin{figure} 
\centering 
       \scalebox{0.6}{\includegraphics{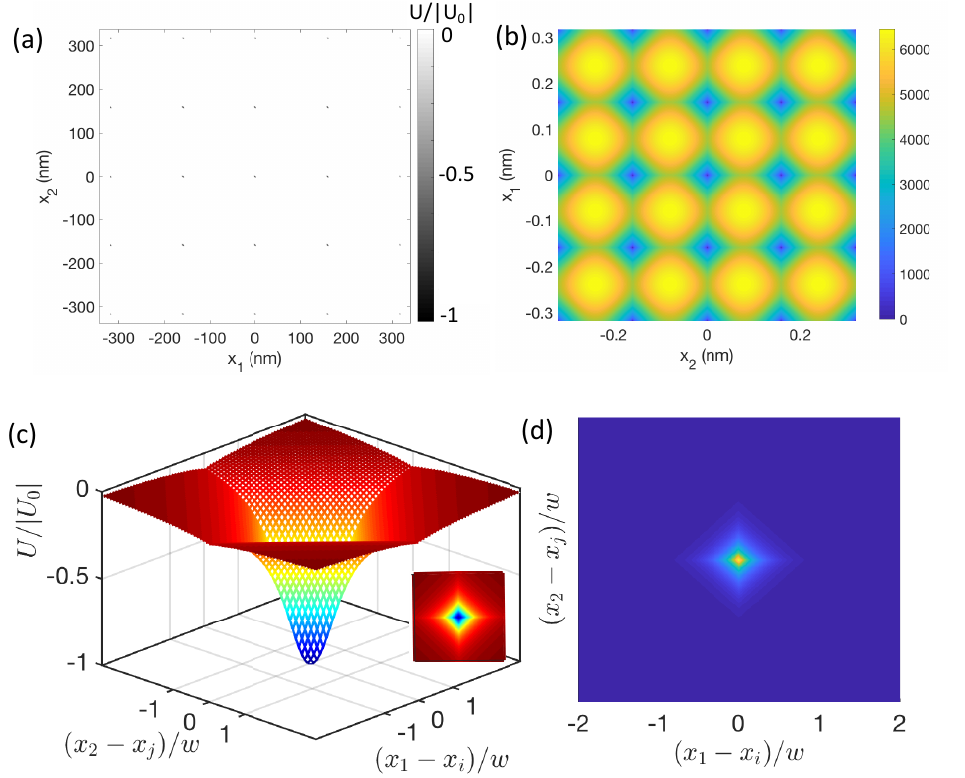}}   
\caption{Rydberg dressed trapping - Spatial correlation. (a) Having both atoms free to move, the trapping potential is numerically evaluated from Eq.~3 as a function of the two atoms' positions $x_1$ and $x_2$. (b) The resonance condition $(\Omega_2(x_1)^2+\Omega_2(x_2)^2-8\Delta^2)/\Gamma_p$ is plotted, which resembles the potential pattern of (a). (c) The zoomed-in picture of trap potential for the case of two atoms being at $x_i$ and $x_j$ nodes shows the correlation between atoms' thermal motion. (inset) The centered diamond equipotential lines force the two atoms to oscillate in phase or out of phase. This phononic bus provides all-to-all connectivity in a multi-dimensional lattice for applying quantum gates via side-band laser driving similar to ion-trap counterparts.
(d) Applying the mean-field approach, the two atoms' ground state wave-function at specific lattice site  $x_i$ and $x_j$ is plotted. Laser parameters are presented in Fig.~2.
 }\label{FigMeanfield}
\end{figure}

The origin of the enhanced interaction can be traced to a two-atom resonance that occurs in the presence of strong interaction \cite{Gau16}.
Considering the laser coupling of two interacting atoms plotted in Fig.~\ref{Fig3},
for $\Omega_1\ll\Omega_2$ the two-atoms Hilbert space would be organized in three subspaces that are coupled by weak $\Omega_1$ laser. These subspaces are the ground state $\ket{gg}$, one atom excitation \{$\ket{gp,pg}$, $\ket{ge,eg}$\}, and two atom excitation states \{$\ket{pp}$, $\ket{pe,ep}$, $\ket{ee}$\}.
The strong coupling $\Omega_2$, mixes the states in each subspace. 
Pre-diagonalizing the subsystems quantifies the light-shifts experienced by the eigen-states, see Fig.~\ref{Fig3}b.
For the second subspace with single excitation in $i^{th}=\{1,2\}$ atom,  the coupling Hamiltonian in the  \{$\ket{gp}$, $\ket{ge}$\} or \{$\ket{pg}$, $\ket{eg}$\} basis is given by  
\begin{equation}
S_2^{i}/\hbar=\begin{pmatrix}
-\Delta & \Omega_2(x_{i})/2\\
\Omega_2(x_{i})/2 & 0
\end{pmatrix}.
\end{equation}
The eigen-energies  in this subspace $\beta_{i_\pm}/\hbar=-\Delta/2\pm1/2\sqrt{\Delta^2+\Omega_2(x_{i})^2}$ does not get resonant with the ground state. 
The coupling Hamiltonian in the third subspace with double excitation basis  \{$\ket{pp}$, $\ket{pe}$, $\ket{ep}$, $\ket{ee}$\}   is given by
\begin{equation}
\frac{S_3}{\hbar}=\begin{pmatrix}
-2\Delta & \Omega_2(x_1)/2 & \Omega_2(x_2)/2&0\\
 \Omega_2(x_1)/2& -\Delta &0 &  \Omega_2(x_1)/2 \\
 \Omega_2(x_2)/2 &0 & -\Delta& \Omega_2(x_2)/2\\
 0&\Omega_2(x_1)/2 &\Omega_2(x_2)/2 & V \\
\end{pmatrix}
\label{Eq_S3}
\end{equation}
For large interaction inside the soft-core $V\rightarrow \infty$,  the doubly excited Rydberg state $\ket{\lambda_0}\approx \ket{ee}$ decouples asymptotically. Hence the eigen-energies would be  $\lambda_{0}=\hbar V$, $\lambda_{1}=-\hbar \Delta$ and $\lambda_{\pm}=-\frac{3\hbar}{2}\Delta\pm\hbar/2\sqrt{\Delta^2+\Omega_2(x_1)^2+\Omega_2(x_2)^2}$.  
At $\Omega_2(x_1)^2+\Omega_2(x_2)^2=8\Delta^2$, the $\ket{ \lambda_{-}}$ eigen-state couples resonantly with the ground state, generating an enhanced light-shift. As discussed above, a small deviation of laser intensity from the resonance condition $\Omega_2(x_1)^2+\Omega_2(x_2)^2-8\Delta^2= \gamma_p^2$ results in a significant change of the trapping interaction. Hence, the interaction-induced resonant peaks would be localized at very specific points of the $\Omega_2(x)$ standing-wave. 
This resonance condition generates a bipartite motional correlation, where the atoms oscillate in phase or out of phase, see Fig. \ref{FigMeanfield}. This would generate a phononic bus that could resemble the atomic version of ion trap gates \cite{Cir95} by driving the side-band transitions.
 
\section{Decoherence}

 \subsection{Spontaneous emission}
The main source of {\it decoherence} in the proposed lattice is the spontaneous emission from the intermediate state. 
 Rydberg interaction disturbs individual atom's dark state, populating the intermediate state $|p\rangle$, and hence increases the loss rate per atom $\Gamma=\text{Tr}[{\rho}_{i} (\gamma_p {\sigma}_{pp}+\gamma_e {\sigma}_{ee})]$ at the trapping potential teeth. 
 The loss rate spatial profile, $\gamma_p U_{i,j}(x_1,x_2)/\Delta$,  follows the interaction profile. This is because, at the trapping points, the doubly excited states in the third subspace presented in Eq.~\ref{Eq_S3} would get in resonance with the laser and enhance the population of the short-lived $p$ states. 
 
  The maximum loss for a given $\Omega_2(x)$ profile could be controlled by adjusting the intensity of $\Omega_1$ laser. The scale of trap depth for two atoms located within the soft-core is plotted in Fig.~\ref{Fig5}a as a function of $\Delta$ for the maximum scattering rate of 1Hz in the pair interaction. The interaction-to-loss ratio is enhanced by applying stronger laser driving of $\Omega_{2}$. Having $N$ single-atom-occupied trapping sites within the $\pm R_c$  interaction distance, the trapping potential experienced by an atom would add up to $NU_0$.

 \subsection{Laser Noise Heating}

To analyze the heating rate caused by laser intensity fluctuations, we begin with the model Hamiltonian for a trapped atom of mass \( m \):
\begin{equation}\label{Eq_trappingH}
H=\frac{P^2}{2m}+\frac{1}{2}m\omega_{tr}^2x^2
\end{equation}
where \( \omega_{tr} \) is the trap frequency, which depends on the laser intensities as \( \omega_{tr}^2 \propto I_{2sw} I_1^2 \) (see Eq.~\ref{Eq_omegatrap}). Fluctuations in the laser intensities, denoted by \( \epsilon_j = \frac{I_j - I_{j0}}{I_{j0}} \), modify the trap frequency, leading to the effective transformation \( \omega_{tr}^2 \rightarrow (1 + \epsilon_{2sw}(t))(1 + \epsilon_1(t))^2 \omega_{tr}^2 \).

The heating rate can be determined by calculating the average transition rates between quantum states of the trap using first-order time-dependent perturbation theory.
To first order in the intensity fluctuations, the perturbation to the Hamiltonian from Eq.~\ref{Eq_trappingH} is expressed as:
\begin{equation}
 H'(t)=\frac{1}{2}(\epsilon_{2sw}(t)+2\epsilon_1(t)) m\omega_{tr}^2x^2.
\end{equation}
For an atom in the motional state \( \ket{n} \), the transition rate to the states \( \ket{n \pm 2} \) due to \( H'(t) \) is given by \cite{Sav97}:
\begin{equation}
\label{Eq_transition}
R_{n \pm 2 \leftarrow n} = \frac{\pi \omega_{tr}^2}{16} \left( 4 S_{\epsilon_1}(2\omega_{tr}) + S_{\epsilon_{2sw}}(2\omega_{tr}) \right) (n+1 \pm 1)(n \pm 1),
\end{equation}
where \( S_{\epsilon} \) is the one-sided power spectrum density of the fractional intensity noise, defined as:
 \begin{equation}
S_{\epsilon}(2\omega_{tr}) \equiv \frac{2}{\pi} \int_0^\infty d\tau \, \cos(2\omega_{tr} \tau) \langle \epsilon(t) \epsilon(t + \tau) \rangle.
\end{equation}

Assuming that at time \( t \), the trapped atoms occupy the state \( \ket{n} \) with probability \( P(n,t) \), the average energy is given by $\langle E(t) \rangle = \sum_n P(n,t) \left( n + \frac{1}{2} \right) \hbar \omega_{tr}$. Considering the transition rates in Eq.~\ref{Eq_transition}, the average energy increases over time as:
 \begin{equation}
\langle \dot{E} \rangle = \Gamma_{\epsilon} \langle E \rangle, \quad \Gamma_{\epsilon} = \pi^2 \nu_{tr}^2 S_{\epsilon}(2\nu_{tr}),
\end{equation}
where $\nu_{tr}$ is the trap oscillation frequency.

For the $\Omega_1$ laser, in a classical strontium trap \cite{Wan20}, the linear power spectrum  over  a 360kHz bandwidth is reported as $S_{\epsilon}(2\nu_{tr})=2.3\times10^{-13}$ Hz$^{-1}$  for the servo-active case, and \( 4 \times 10^{-12} \, \text{Hz}^{-1} \) for the free-running case.
 Using these values for the noise spectra, the corresponding heating rates are \( \Gamma_{\text{free}} = 5.1 \, \text{s}^{-1} \) and \( \Gamma_{\text{servo}} = 0.3 \, \text{s}^{-1} \).
   In the case of the $\Omega_2 $ laser, a 362 nm laser has been used to trap \(^{199}\text{Hg}\) atoms in an optical lattice with a trap frequency of 128 kHz \cite{Mej11}. Following the approach of \cite{Sav97}, they calculated the heating rate due to laser intensity noise, which results in a trap lifetime on the order of several seconds.

 Note that at large frequencies $\nu_{tr}>25$kHz, the power spectrum scales as $\nu_{tr}^{-2}$ \cite{Sav97,Jia23}.
In the proposed optical trap, as the trap frequency increases, the contribution of laser intensity noise to heating diminishes due to this scaling. However, the overall heating rate stabilizes because of a balance between the increasing trap frequency and the decreasing noise spectrum. 
As a result, the decay rate remains nearly constant for trap frequencies above 25 kHz, as shown in Fig. 1b of \cite{Sav97}. In conclusion, the above experimental reports of the laser noise-induced heating rate are expected to be valid for the range of trap frequencies that are studied in the proposed quantum trap. 

\subsection{Sideband Heating}
Finally, laser manipulation of the atomic state can introduce position-dependent phases, potentially causing heating through sideband excitation of higher motional states. In the trap discussed in Fig.~\ref{Fig5}c, the Lamb-Dicke parameter is $\eta=k_1\sigma/\sqrt{2}=0.025$, considering the wavefunction half-width at $1/e$ maximum of $\sigma=4$ nm interacting with the 689 nm laser. For the sideband transition, the effective two-photon Rabi frequency between ground and Rydberg state, \( \eta \Omega_1 \Omega_2 / 2\Delta \), is five orders of magnitude smaller than the trap frequency, \( \omega_{tr} \), which corresponds to the detuning from the excited motional states. As a result, the probability of exciting higher motional states—and the associated heating—is negligible.
  
\section{Realization}  
 
  \begin{figure} [h]
\centering 
       \scalebox{0.46}{\includegraphics{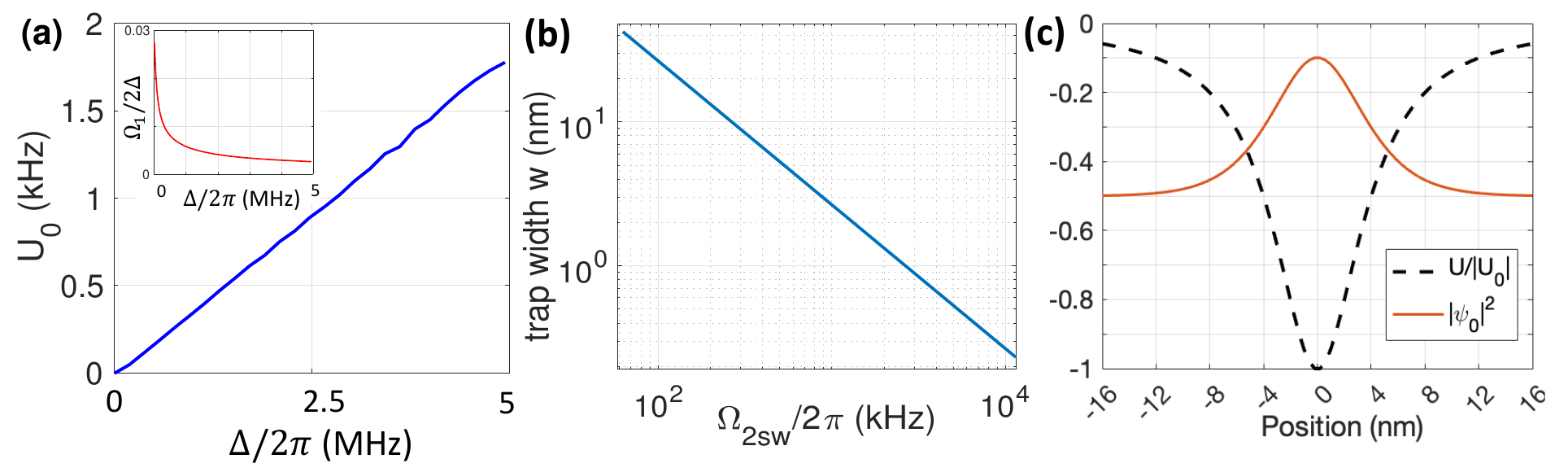}} 
\caption{{\bf (a)} The scale of trap depth $U_0$ for two atoms located within the soft-core distance of $R_c$ is plotted as a function of $\Delta=\Omega_{2max}/2$ for the constant scattering rate of 1Hz. Having $N$ lattice sites within the interaction distance $R_c$, the trapping potential experienced by an atom would add up to $NU_0$.
 (inset) The decoherence rate is adjusted to 1Hz by controlling the ratio of $\Omega_1/2\Delta$.  {\bf (b)} The width of Lorentzian traps $w$ (Eq.~\ref{Eq_Width}) is plotted as a function of  $\Omega_{2sw}$ for $\theta=\pi$. This is in contrast with the classical trap width of $\lambda/4$.   Here $w/\lambda_2=\{0.1,0.01,0.001\}$ for $\Omega_{2sw}/2\pi=\{0.09,\, 0.9,\, 9\}$MHz respectivly. {\bf (c)} The trapping potential and ground eigenstate are numerically obtained for a 3D lattice formed by dressing atoms to $|100S\rangle$  with $\Delta/2\pi=5$MHz, $\Omega_{2sw}/2\pi=650$kHz. and $\Omega_{2max}/2\pi=10$MHz. The collective interaction adds up to a single site trap depth of $U_0=37$MHz with a trap width of $w=4$nm. }\label{Fig5}
\end{figure}

   Figure~\ref{Fig5} plots the trap width and depth in a pair interaction as a function of applied laser parameters. The loss rate per atom is controlled to 1Hz by adjusting the $\Omega_1$. The effective trap depth would be the sum of all pair interactions encountering neighboring sites within the $R_c$ distance.
For example, dressing ground state atoms to $\ket{5s100s\, ^3S_1}$ with $\Omega_{2max}/2\pi=10$MHz and limiting the loss rate per atom in an interacting pair to 1Hz, the collective trapping potential experienced by a single atom in 3D lattice interacting by neighboring sites within the soft-core would be $N_{3D}U_0$=37MHz. 
For the Lorentzian trap of Eq.~\ref{Eq_Lorentzian}, to ensure that the potential well is deep enough to support at least one bound state, the depth $U_0$ and width $w$ must be related by $U_0 w^2 > \frac{\hbar^2}{2m}$. Hence in the mentioned 3D lattice, atom confinement in ultra narrow traps with Lorentzian width of $w>3$nm is possible.
 
 The ground motional state of the mentioned trap is numerically obtained by solving the Schrodinger equation
    \begin{equation}
    -\frac{\hbar^2}{2m} \frac{d^2 \psi(x)}{dx^2} + \frac{U_0}{1 + \left(\frac{x}{w}\right)^2} \psi(x) = E\psi(x),
  \end{equation}    
  by discretizing the spatial coordinate, using the finite difference method, and solving for the eigenvalues \(E\) and eigenfunctions \(\psi(x)\).  The only trapped eigenstate is plotted in Fig. \ref{Fig5}c for a case with $U_0=37$MHz and $w=4$nm. The atom is confined to a HWHM of 4nm.

The quantum lattice improves the atom confinement to a level that is not achievable for classical optical lattices with the price of using an extra laser. In the case of strontium, the $\Omega_2$ is in the UV spectrum, where strong power would be more expensive compared to the optical lasers. The importance of the UV laser's power can be seen in  Eq.~\ref{Eq_Width} where $\Omega_{2sw}$ and $\Omega_{2max}=2|\Delta|$ are contributing to the width and the depth of the trap. 
The available UV lasers in the range of 316.3nm-319.3nm with powers greater than 200mW \cite{Bri16} and 300mW \cite{Toptica} allow exciting the range of Rydberg states between $n=35$ to $n>300$. For making a 2D standing wave with homogeneous plateau-type laser intensity, a super-Gaussian focusing profile of $\exp(-x^{10}/\sigma_x^{10})\exp(-y^{10}/\sigma_y^{10})$ with $\sigma_x=10\mu$m and $\sigma_y=400\mu$m is considered. The 200mW power generates an effective Rabi frequency of $\Omega_{2}/2\pi=20$MHz for addressing $\ket{5s100s \, ^3S_1}$ Rydberg level. 
In the case of 3D lattices using lower principal numbers is recommended to enhance the transition dipole moment which scales by $n^{-3/2}$. Note that the blockage radius could be as a small as lattice constant allowing a significant reduction of $n$.
In the case of Alkaline atoms, the transitions are controlled by optical lasers where stronger power is available. For example in cesium, the optimum intermediate states are $7P$ and $5D$ with $\gamma=165$ and 1353ns that are reachable by electric dipole and quadrupole respectively.

\section{Quantum vs Classical Traps - Advantages}

In comparison with the conventional AC stark shift lattices, the proposed quantum lattice is superior in terms of atom confinement. To comapre the two schemes we consider a regime where the trapping potential in both cases could be approximated by a harmonic trap. The Hamiltonian of a particle inside the trap would then be given by Eq.~\ref{Eq_trappingH}. The motional states of an atom in these harmonic traps are 
\begin{equation}
\phi_n(x)=\sqrt{\frac{1}{\sqrt{\pi}2^n n! \sigma_0}} e^{-\frac{1}{2}(\frac{x}{\sigma_0})^2}H_n(x/\sigma_0)
\end{equation}
where $H_n$ is the Hermit polynomial and $\sigma=\sqrt{\hbar/m\omega_t}$ is the size of motional ground state $\phi_0(x)$.
In classical optical lattices, the trap potential is exclusively provided by $\Omega_1$ laser which is given by $U^{cl}(x)=2U^{cl}_0\cos^2(k_1x)$ with $U^{cl}_0=-1/2\alpha \mathcal{E}^2$ where  $\mathcal{E}^2$ is the intensity of the trapping field. The potential of a single site could be approximated by the Harmonic model e.g.   $2|U^{cl}_0|(-1+k_1^2(x-x_j)^2)$ for the  $j^{th}$ trap site.  Hence the classical trap frequency would be given by $\omega_t^{cl}=k_1\sqrt{2U_0/m}$. Corresponding  atom confinement  
\begin{equation}
\sigma^{cl}=\sqrt{\frac{\hbar}{k_1\sqrt{2U_0m}}}=\sqrt[4]{\frac{\hbar \Delta}{m}}\sqrt{\frac{2}{k_1 \Omega_1}}\propto(\mathcal{E}^{2})^{-1/4}\lambda_1^{1/2}
\end{equation}
 is inversely proportional to the quadruple root of the laser intensity.

\begin{figure} [h]
\centering 
       \scalebox{0.5}{\includegraphics{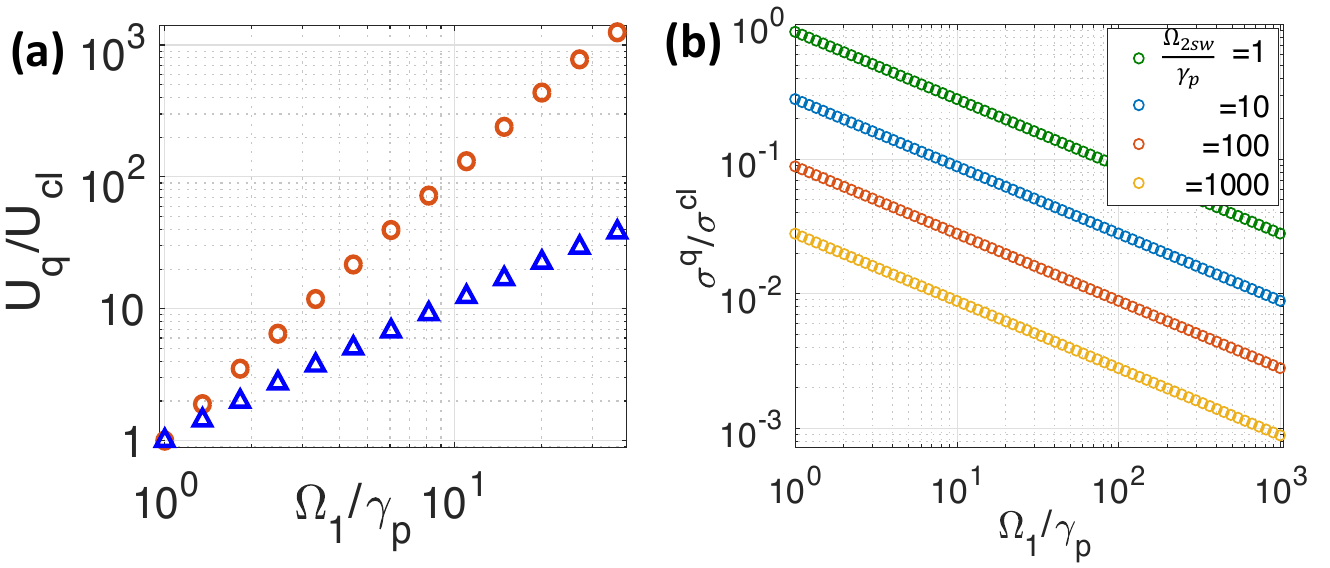}} 
\caption{{\bf Quantum vs. Classical Trap Comparison. (a)} 
The plot shows the ratio of the quantum to classical trap strengths, $ U_q/U_{cl} $, as a function of $ \Omega_1 $.  
Circle markers indicate the scenario where the detuning is the same for both traps, \( \Delta_q = \Delta_{cl} \), and the decay rate is fixed at \( \Gamma = 1 \, \text{Hz} \) for the quantum trap. 
Triangle markers represent the case where the detunings are adjusted independently, \( \Delta_q \neq \Delta_{cl} \), such that \( \Gamma = 1 \, \text{Hz} \) is maintained in both the quantum and classical traps.  
 {\bf (b)} Relative confinement efficiency in terms of laser power. 
The Gaussian width of the ground motional state is compared between the quantum and classical traps, with the ratio given by $\frac{\sigma_q}{\sigma_{cl}}=\sqrt{2\frac{k_1}{k_2}\frac{\gamma_p}{\Omega_{2sw}} \frac{\gamma_p}{\Omega_1}}$.
}\label{Fig6}
\end{figure}

In quantum lattice, the Lorentzian trap potential of Eq.~\ref{Eq_Lorentzian} can be approximated by a Harmonic oscillator potential  over small spatial range $x-x_i\ll w$ and with  $\Delta<0$ as
\begin{equation}
U^q=|U^q_0|(-1+\frac{(x-x_i)^2}{w^2})= \frac{\hbar\Omega_1^4}{8|\Delta|\gamma_p^2}(-1+\frac{k_2^2\Omega_{2sw}^2}{4\gamma_p^2}(x-x_i)^2)
\end{equation}
where trap depth and width are defined in Eq.~\ref{Eq_Width} and $k_2$ is the wave-vector of  $\Omega_2$ laser. The trap frequency and atom confinement in the quantum lattice are given by 
\begin{eqnarray}
\label{Eq_omegatrap}
\omega_{tr}^q=\frac{k_2\Omega_{2sw}}{2\gamma_p}\sqrt{\frac{2U^q_0}{m}}, \quad
\sigma_q=\frac{2\gamma_p}{\Omega_1\sqrt{k_2 \Omega_{2sw}}} \sqrt[4]{\frac{\hbar \Delta}{ m}}\quad \quad
\end{eqnarray}
Figure \ref{Fig6}b compares the atom confinement in quantum and classical traps as a function of laser power. It clearly shows a cost-benefit advantage of the quantum model for ultra-tight localization of atoms in the lattice. 
The quantum trap greatly amplifies the restoring force, which is determined by the potential's gradient. This is obtained by increasing the potential depth while compressing the potential width, see Fig.~\ref{Fig5}a,b.
 While the potential width in classical traps is limited to the laser wavelength, in quantum traps the width is suppressed to the deep sub-wavelength limit for $\gamma_p\ll\Omega_{2sw}$, see Eq.~\ref{Eq_Width}.  So while the width of the strontium classical trap is defined by a quarter of magic wavelength $\lambda_1/4= 203.75nm$, the width of the quantum trap is one and two orders of magnitude smaller than the classical trap for $\Omega_{2sw}/2\pi=150$kHz  and 1.5MHz respectively, see Fig.~\ref{Fig5}b.
In terms of trap depth, in classical strontium lattices operating with the standing wave $\Omega_1$ clock transition \cite{Tak05,Bil17}, the potential is given by $U^{cl}(r)=\Omega_1^2/8\Delta$. 
The potential depth in the quantum lattice as provided in Eq.~\ref{Eq_Width}, is proportional to the square of the laser intensity i.e. $U^{q}_0=\frac{\hbar\Omega_1^4}{8\Delta\gamma_p^2}$, which provides a deeper potential trap for the case $\Omega_1>\gamma_p$. This means that for $\Omega_1/2\pi=150$kHz and 1.5MHz the quantum trap of only two interacting atoms is $4\times10^4$ and $4\times10^6$ times deeper than the classical counterpart and importantly this ratio scales by the number of atoms in the $R_c$ region. The interaction to the light scattering ratio $U/\Gamma_{sc}=\Delta/\gamma_p$ follows the same relation in both quantum and classical lattice models.

 {\bf Outlook:} Since the effective trapping potential is interaction-induced, there can be motional coupling analogous to phononic excitation in real materials. This could be seen in the correlation of Fig.~\ref{FigMeanfield}c. The motional correlation forms a phononic bus for quantum computation with all-to-all connectivity similar to the ion trap with the major advantage of not being restricted to one dimension lattice. 
 
  The sub-nanoscale resolution in trapping atoms demonstrated here extends the toolbox of neutral atom quantum technology.  Ultra-narrow wells in this proposal allow significant suppression of the lattice constant with a time-sharing approach \cite{Nas15}. In this approach, the applied standing wave is stroboscopically shifted in space by $\lambda/2N$ and hence the effective lattice constant would be smaller by a factor of $N^{-1}$. 
This compaction of the atomic lattice is demanding for scaling the lattice sites with the currently limited laser powers. Furthermore, in quantum simulation with optical lattices, the energy scale of Hubbard models for both hopping and interaction of atoms is set by the minimum lattice constant which used to be limited to $\lambda/2$, leading to challenging temperature requirements to observe quantum phases of interest \cite{Gro17}.
Furthermore, by adjusting $\Omega_{2c}$ one can move the resonance position over the standing wave, which allows bringing pairs of atoms to a very short interatomic distance, which enables controlled ultracold quantum chemistry \cite{Luh15,Osp10,Liu18}.

A distinct research avenue looks at the applications of the presented scheme with ultra-narrow repulsive peaks.  Equation \ref{Eq_5} shows that changing the detuning sign would preserve the interaction profile but only flip the potential sign from attractive to repulsive.  
These ultra-narrow barriers are ideal for realizing the Kronig-Penney (KP) lattice model \cite{Kro93}.  Furthermore, the three-dimensional repulsive $\delta$-function peaks form nearly perfect box-traps \cite{Gau13}. These repulsive narrow peaks also realize thin tunnel junctions for atomtronic devices \cite{Sea07,Eck14}.
The potential is easily generalizable to other three-dimensional geometries using the holographically designed laser intensity \cite{Bar18}.

\section{Appendix II: Rydberg molecule}

The minimum lattice spacing of 160 nm, adjustable by the angle of the counterpropagating beams, may raise concerns about potential molecule formation. 
However, Rydberg molecule formation occurs only when the attractive potential due to Rydberg electron-neutral atom scattering brings two atoms into extremely close proximity (on the order of 2 nm). 
At this distance, the binding energy could ionize the Rydberg electron, forming a Sr\(_2^+\) molecule \cite{Nie15}. 
In the absence of mass transport, stepwise decay or ionization of the Rydberg atom is highly improbable, as the quantization of Rydberg states prevents such processes. This has been theoretically discussed and experimentally verified in \cite{Bal13}. Even at high atomic densities, the molecular binding energy between nearby atoms is orders of magnitude smaller than the spacing between Rydberg levels for all principal quantum numbers, making ion-pair formation highly unlikely \cite{Nie15}. In conclusion, the confinement of atoms in the optical lattice prevents the described mass transport, effectively eliminating the channel for molecule formation losses.


\begin{thebibliography}{99}




\bibitem{Kaz21}M. Khazali, W. Lechner, Scalable quantum processors empowered by the Fermi scattering of Rydberg electrons.\href{https://doi.org/10.1038/s42005-023-01174-4} { {\em Comms. Phys.} {\bf 6}, 57 (2023). }

\bibitem{Khaz22Log} M. Khazali, Universal terminal for cloud quantum computing. \href{https://doi.org/10.1038/s41598-024-65899-0}{{\em Scientific Reports} {\bf 14}, 15412 (2024).}

\bibitem{Kha22} M. Khazali, Discrete-Time Quantum-Walk \& Floquet Topological Insulators via Distance-Selective Rydberg-Interaction,\href{https://doi.org/10.22331/q-2022-03-03-664}{ {\em Quantum} {\bf 6}, 664 (2022). }

\bibitem{Hol22}Hollerith, S., {\it et al.}, Realizing distance-selective interactions in a Rydberg-dressed atom array. \href{https://doi.org/10.1103/PhysRevLett.128.113602}{{\em Phys. Rev. Lett.} {\bf 128}, 113602 (2022).}

\bibitem{Ate07} C. Ates, T. Pohl, T. Pattard, and J. M. Rost,  Antiblockade in Rydberg excitation of an ultracold lattice gas. \href{ https://doi.org/10.1103/PhysRevLett.98.023002}{{\em Phys. Rev. Lett.} {\bf 98}, 023002 (2007).}

\bibitem{Wu22} Wu, J. L., Wang, Y., Han, J. X., Su, S. L., Xia, Y., Jiang, Y. \& Song, J. Unselective ground-state blockade of Rydberg atoms for implementing quantum gates. \href{https://doi.org/10.1007/s11467-021-1104-7}{{\it Front. Phys.} {\bf 17}, 22501 (2022).}


\bibitem{Gra19}Graham, T.M., Kwon, M., Grinkemeyer, B., Marra, Z., Jiang, X., Lichtman, M.T., Sun, Y., Ebert, M. and Saffman, M., Rydberg-mediated entanglement in a two-dimensional neutral atom qubit array. \href{https://doi.org/10.1103/PhysRevLett.123.230501}{{\em Phys. Rev. lett.} {\bf 123}, 230501 (2019).}

\bibitem{Pag22}A. Pagano, S. Weber, D. Jaschke, T. Pfau, F. Meinert, S. Montangero, and H. P. B\"uchler, Error budgeting for a controlled-phase gate with strontium-88 Rydberg atoms, \href{https://doi.org/10.1103/PhysRevResearch.4.033019}{{\em Phys. Rev. Research} {\bf 4}, 033019 (2022).}

\bibitem{Cet22}Cetina, M., Egan, L.N., Noel, C., Goldman, M.L., Biswas, D., Risinger, A.R., Zhu, D. and Monroe, C., Control of transverse motion for quantum gates on individually addressed atomic qubits. \href{https://doi.org/10.1103/PRXQuantum.3.010334}{{\em PRX Quantum} {\bf 3}, 010334 (2022).}


\bibitem{Bre99} Brennen, G.K., Caves, C.M., Jessen, P.S. and Deutsch, I.H.,  Quantum logic gates in optical lattices. \href{https://doi.org/10.1103/PhysRevLett.82.1060}{{\em Phys. Rev. Lett.} {\bf 82}, 1060 (1999).}


 \bibitem{Cir95}Cirac, Juan I., and Peter Zoller. Quantum computations with cold trapped ions. \href{ https://doi.org/10.1103/PhysRevLett.74.4091}{{\em Phys. Rev. Lett.} {\bf 74}(20), 4091 (1995).}

\bibitem{Wan18}Wang, Yang, et al. Dark state optical lattice with a subwavelength spatial structure. \href{https://doi.org/10.1103/PhysRevLett.120.083601}{{\em Phys. Rev. Lett.} {\bf 120},  083601 (2018).}
\bibitem{Tsu20}Tsui, T. C., Wang, Y., Subhankar, S., Porto, J. V., \& Rolston, S. L.  Realization of a stroboscopic optical lattice for cold atoms with subwavelength spacing. \href{https://doi.org/10.1103/PhysRevA.101.041603}{{\em Phys. Rev. A}, {\bf 101}, 041603 (2020).}

\bibitem{Saf10}M. Saffman, T. G. Walker, and K. M\o lmer. Quantum information with Rydberg atoms. \href{https://doi.org/10.1103/RevModPhys.82.2313}{{\em Rev. Pod. Phys.} {\bf 82}, 2313 (2010).}

\bibitem{Kha20}M. Khazali and K. M{\o}lmer.
\newblock Fast multiqubit gates by adiabatic evolution in interacting excited-state manifolds of Rydberg atoms and superconducting circuits.
\href{https://doi.org/10.1103/PhysRevX.10.021054}{ {\em Phys. Rev. X} {\bf 10}, 021054, (2020).}

\bibitem{Kha19}M. Khazali, C.~R Murray, and T. Pohl.
\newblock Polariton exchange interactions in multichannel optical networks.
\href{ https://doi.org/10.1103/PhysRevLett.123.113605}{ {\em Phys. Rev. Lett.} {\bf 123} 113605, (2019).}

\bibitem{Kha15}M. Khazali, K. Heshami, and C. Simon.
\newblock Photon-photon gate via the interaction between two collective Rydberg
  excitations.
\href{https://doi.org/10.1103/PhysRevA.91.030301}{ {\em Physical Review A}, {\bf 91} 030301, (2015).}

\bibitem{Khaz22AllOptic} M. Khazali, All-optical quantum information processing via a single-step Rydberg blockade gate. \href{https://doi.org/10.1364/OE.481256}{{\em  Optics Express} {\bf 31}(9), 13970-13980 (2023). }

\bibitem{Kha17}M. Khazali, K. Heshami, and C. Simon.
\newblock Single-photon source based on Rydberg exciton blockade.
\href{https://doi.org/10.1088/1361-6455/aa8d7c}{ {\em J. Phys. B: At. Mol. Opt. Phys.}
  {\bf 50}, 215301, (2017).}
 


\bibitem{KazRev}M. Khazali,  "Quantum information and computation with Rydberg atoms." \href{https://doi.org/10.22051/ijap.202134445.1188}{{\em Iranian Journal of Applied Physics} {\bf 10} 19 (2021)}; M. Khazali, Applications of Atomic Ensembles for Photonic Quantum Information Processing and Fundamental Tests of Quantum Physics. Diss. University of Calgary (Canada), (2016).

\bibitem{Zei16} J. Zeiher, R. Van Bijnen, P. Schaus, S. Hild, J. Choi, T. Pohl, I. Bloch, and C. Gross. Many-body interferometry of a Rydberg-dressed spin lattice. \href{https://doi.org/10.1038/nphys3835}{{\em Nature Physics} {\bf 12}, 1095-1099 (2016).}
\bibitem{Hin23} Hines, J. A., Rajagopal, S. V., Moreau, G. L., Wahrman, M. D., Lewis, N. A., Markovi, O., \& Schleier-Smith, M. Spin Squeezing by Rydberg Dressing in an Array of Atomic Ensembles. \href{https://doi.org/10.1103/PhysRevLett.131.063401}{{\em Phys. Rev. Lett.} {\bf 131}, 063401 (2023).}

\bibitem{Eck23} Eckner, William J., Nelson Darkwah Oppong, Alec Cao, Aaron W. Young, William R. Milner, John M. Robinson, Jun Ye, and Adam M. Kaufman. Realizing spin squeezing with Rydberg interactions in a programmable optical clock.  \href{https://doi.org/10.1038/s41586-023-06360-6}{{\em Nature} {\bf 621}, 734-739 (2023).}

\bibitem{Kha16}M. Khazali, H.~W. Lau, A. Humeniuk, and C. Simon.
\newblock Large energy superpositions via Rydberg dressing.
\href{https://doi.org/10.1103/PhysRevA.94.023408}{ {\em Phys. Rev. A} {\bf 94}, 023408, (2016)}.
\bibitem{Kha18}M. Khazali,
\newblock Progress towards macroscopic spin and mechanical superposition via  Rydberg interaction.
\href{https://doi.org/10.1103/PhysRevA.98.043836}{ {\em Phys. Rev. A} {\bf 98}, 043836, (2018).}
\bibitem{Khazal24}M. Khazali, Fast multicomponent cat-state generation under resonant or strong-dressing Rydberg-Kerr interaction, \href{https://doi.org/10.1103/PhysRevA.109.053716}{{\em Phys. Rev. A} {\bf 109}, 053716 (2024).}

\bibitem{San00}Santos, L., Shlyapnikov, G. V., Zoller, P., \& Lewenstein, M.  Bose-Einstein condensation in trapped dipolar gases. \href{https://doi.org/10.1103/PhysRevLett.85.1791}{{\em Phys. Rev. Lett.} {\bf 85}, 1791 (2000).}
\bibitem{Hon10}Honer, J., Weimer, H., Pfau, T., \& B\"uchler, H. P. Collective many-body interaction in Rydberg dressed atoms. \href{https://doi.org/10.1103/PhysRevLett.105.160404}{{\em Phys. Rev. Lett.} {\bf 105}, 160404 (2010).}

\bibitem{Gau16}C. Gaul, B. J. DeSalvo, J. A. Aman, F. B. Dunning, T. C. Killian, and T. Pohl, Resonant Rydberg Dressing of Alkaline-Earth Atoms via Electromagnetically Induced Transparency, \href{ https://doi.org/10.1103/PhysRevLett.116.243001}{{\em Phys. Rev. Lett.} {\bf 116}, 243001 (2016).}
\bibitem{Kha21}M. Khazali, Rydberg noisy dressing and applications in making soliton molecules and droplet quasicrystals, \href{https://doi.org/10.1103/PhysRevResearch.3.L032033}{{\em Phys. Rev. Research} {\bf 3}, L032033 (2021).}

\bibitem{Hen12}Henkel, N., Cinti, F., Jain, P., Pupillo, G. and Pohl, T.,  Supersolid vortex crystals in Rydberg-dressed Bose-Einstein condensates. \href{https://doi.org/10.1103/PhysRevLett.108.265301}{{\em Phys. Rev. Lett.} {\bf 108}, 265301 (2012).}


\bibitem{Shi21}Shi, Zeyun, and Guoxiang Huang. Self-organized structures of two-component laser fields and their active control in a cold Rydberg atomic gas. \href{https://doi.org/10.1103/PhysRevA.104.013511}{{\it Phys. Rev. A} {\bf 104}, 013511 (2021).}

\bibitem{Shi20}Shi, Zeyun, Weibin Li, and Guoxiang Huang. Structural phase transitions of optical patterns in atomic gases with microwave-controlled Rydberg interactions. \href{https://doi.org/10.1103/PhysRevA.102.023519}{{\em Phys. Rev. A} {\bf 102}, 023519 (2020).}

\bibitem{ZShi21}Z. Shi, and G. Huang, Selection and cloning of periodic optical patterns with a cold Rydberg atomic gas. \href{ https://doi.org/10.1103/PhysRevA.102.023519}{{\em Optics Letters} {\bf 46}, 5344-5347 (2021).}

\bibitem{Shi23}Shi, Zeyun, Fazal Badshah, and Lu Qin. Two-dimensional lattice soliton and pattern formation in a cold Rydberg atomic gas with nonlocal self-defocusing Kerr nonlinearity. \href{https://doi.org/10.1016/j.chaos.2022.112886}{{\em Chaos, Solitons \& Fractals} {\bf 166},  112886 (2023).}
\bibitem{ZShi23}Shi, Zeyun, et al. "Faraday pattern formations in temporally driven Rydberg-dressed Bose-Einstein condensates." \href{https://doi.org/10.1103/PhysRevA.108.063317}{{\em Phys. Rev. A} {\bf 108}, 063317 (2023).}
\bibitem{ZeyShi23}Shi, Zeyun, et al. Spatially modulated control of pattern formation in a general nonlocal nonlinear system. \href{https://doi.org/10.1016/j.chaos.2023.113929}{{\em Chaos, Solitons \& Fractals} {\bf 175}, 113929(2023).}
\bibitem{Shi22}Shi, Zeyun, et al. "Optical pattern formation in a rydberg-dressed atomic gas with non-hermitian potentials. \href{https://doi.org/10.3390/photonics9110856}{{\em Photonics} {\bf 9}, 11 (2022).}
\bibitem{Shi24}Shi, Z., Khazali, M., Qin, L., Zhou, Y., \& Zhong, Y. Pattern formations and their active manipulation in a Rydberg noisy-dressed Bose–Einstein condensate. \href{https://doi.org/10.1364/OL.536991}{{\em Optics Letters} {\bf 49} (2024): 6517-6520.}
\bibitem{ZShi24}Shi, Zeyun, {\it et al.}, Optical pattern formation of laser fields in the Rydberg atomic gases. {\em Optics Express} {\bf 32} (2024): 35366-35380.
\bibitem{ZeyShi24}Shi, Zeyun, et al. Optical solitons and optical patterns controlled by a moiré lattice potential in a Rydberg atomic gas. \href{https://doi.org/10.1103/PhysRevA.110.023513}{{\em Phys. Rev. A} {\bf 110}, 023513 (2024).}


\bibitem{Kun93}S. Kunze, R. Hohmann, H. J. Kluge, J. Lantzsch, L. Monz, J. Stenner, K. Stratmann, K. Wendt, and K. Zimmer, Lifetime measurements of highly excited Rydberg states of strontium I, \href{https://doi.org/10.1007/BF01426757}{{\em Z. Phys. D} {\bf 27}, 111 (1993).}

\bibitem{Sch01}N. Schlosser, G. Reymond, I. Protsenko, and P. Grangier, Sub-poissonian loading of single atoms in a microscopic dipole trap.
\href{ https://doi.org/10.1038/35082512}{ {\em Nature } {\bf 411}, 1024 (2001). }

\bibitem{Sch02}N. Schlosser, G. Reymond, and P. Grangier, Collisional blockade in microscopic optical dipole traps.
\href{https://doi.org/10.1103/PhysRevLett.89.023005}{{\em Phys. Rev. Lett.} {\bf 89}, 023005 (2002).}

\bibitem{Sav97} Savard, T. A., O’hara, K. M., \& Thomas, J. E., Laser-noise-induced heating in far-off resonance optical traps. \href{https://doi.org/10.1103/PhysRevA.56.R1095}{{\em Phys. Rev. A}, {\bf 56}, R1095 (1997).}



\bibitem{Wan20} Wang, Y., Wang, K., Fenton, E. F., Lin, Y. W., Ni, K. K., \& Hood, J. D.  Reduction of laser intensity noise over 1 MHz band for single atom trapping. \href{https://doi.org/10.1364/OE.405002}{{\em Optics Express}, {\bf 28}, 31209 (2020).}

\bibitem{Mej11} Mejri, S., Mcferran, J. J., Yi, L., Le Coq, Y., \& Bize, S.  Ultraviolet laser spectroscopy of neutral mercury in a one-dimensional optical lattice. \href{https://doi.org/10.1103/PhysRevA.84.032507}{{\em Phys.Rev. A} {\bf 84}, 032507 (2011).}

\bibitem{Jia23}Jiang, X., Scott, J., Friesen, M., \& Saffman, M.  Sensitivity of quantum gate fidelity to laser phase and intensity noise. \href{https://doi.org/10.1103/PhysRevA.107.042611}{{\em Phys. Rev. A}, {\bf 107}, 042611 (2023).}

\bibitem{Bri16}Bridge, E. M., Keegan, N. C., Bounds, A. D., Boddy, D., Sadler, D. P., \& Jones, M. P. Tunable cw UV laser with $<$35kHz absolute frequency instability for precision spectroscopy of Sr Rydberg states. \href{https://doi.org/10.1364/OE.24.002281}{{\em Optics Express} {\bf 24}, 2281 (2016).}


\bibitem{Toptica} \url{https://www.toptica.com/fileadmin/Editors_English/11_brochures_datasheets/01_brochures/toptica-br-rydberg_lo.pdf}

\bibitem{Tak05}M. Takamoto, F.-L. Hong, R. Higashi, and H. Katori. An optical lattice clock. \href{https://doi.org/10.1038/nature03541}{{\em Nature} {\bf 435}, 321 (2005).}

\bibitem{Bil17} S. Bilicki. Strontium optical lattice clocks : clock comparisons for timescales and fundamental physics applications. Physics [physics]. \href{https://theses.hal.science/tel-01691598/}{Universite Pierre et Marie Curie - Paris VI, 2017.}
\bibitem{Nas15}S. Nascimbene, N. Goldman, N. R. Cooper, and J. Dalibard, Dynamic optical lattices of sub-wavelength spacing for ultracold atoms, \href{ https://doi.org/10.1103/PhysRevLett.115.140401}{{\em Phys. Rev. Lett.} {\bf 115}, 140401 (2015).}

\bibitem{Gro17} C. Gross and I. Bloch, Quantum simulations with ultracold atoms in optical lattices. \href{https://doi.org/10.1126/science.aal3837}{{\it Science} {\bf 357}, 995 (2017).}



\bibitem{Luh15} Lühmann D-S, Weitenberg C, and Sengstock K, Emulating molecular orbitals and electronic dynamics with ultracold atoms. \href{https://doi.org/10.1103/PhysRevX.5.031016}{ {\em Phys. Rev. X} {\bf 5}, 031016 (2015).}

\bibitem{Osp10} Ospelkaus S, {\it et al.} 
Quantum-state controlled chemical reactions of ultracold potassium-rubidium molecules. \href{https://doi.org/10.1126/science.1184121}{{\em Science} {\bf 327}, 853 (2010).} 
\bibitem{Liu18}Liu LR, Hood JD, Yu Y, Zhang JT, Hutzler NR, Rosenband T, and Ni KK, Building one molecule from a reservoir of two atoms. \href{https://doi.org/0.1126/science.aar7797}{{\em Science} {\bf 360}, 900
(2018). }

\bibitem{Kro93}R. de L. Kronig and W. G. Penney, Quantum mechanics of electrons in crystal lattices.
\href{https://doi.org/10.1098/rspa.1931.0019}{{\em  Proc. Roy. Soc. A} {\bf 130}, 499 (1931).}

\bibitem{Gau13}A. L. Gaunt, T. F. Schmidutz, I. Gotlibovych, R. P. Smith, and Z. Hadzibabic, Bose-Einstein condensation of atoms in a uniform potential, \href{https://doi.org/10.1103/PhysRevLett.110.200406}{{\em  Phys. Rev. Lett.} {\bf 110}, 200406 (2013).}

\bibitem{Sea07}BT. Seaman, M. Kr\"amer, DZ. Anderson, MJ. Holland, Atomtronics: Ultracold-atom analogs of electronic devices. \href{https://doi.org/10.1103/PhysRevA.75.023615}{{\em Phys. Rev. A} {\bf 75},  023615 (2007).}


\bibitem{Eck14} S. Eckel, J. G. Lee, F. Jendrzejewski, N. Murray, C. W. Clark, C. J. Lobb, W. D. Phillips, M. Edwards, and G. K. Campbell, Hysteresis in a quantized superfluid atomtronic circuit. \href{https://doi.org/10.1038/nature12958}{{\em  Nature} {\bf 506}, 200 (2014).}

\bibitem{Bar18}Barredo, D., Lienhard, V., De Leseleuc, S., Lahaye, T. and Browaeys, A., Synthetic three-dimensional atomic structures assembled atom by atom. \href{https://doi.org/10.1038/s41586-018-0450-2}{{\em  Nature} {\bf 561}, 79 (2018).}

\bibitem{Nie15} T. Niederprum, O. Thomas, T. Manthey, T. M. Weber, and H. Ott, \href{https://doi.org/10.1103/PhysRevLett.115.013003}{{\em Phys. Rev. Lett.} {\bf 115}, 013003 (2015).}

\bibitem{Bal13} J. B. Balewski, A. T. Krupp, A. Gaj, D. Peter, H. P. B\"uchler, R. L\"ow, S. Hofferberth, and T. Pfau, Coupling a single electron to a Bose–Einstein condensate. \href{https://doi.org/10.1038/nature12592}{{\em Nature} (London) {\bf 502}, 664 (2013).}


\end{thebibliography}
\end{document}